\documentclass[%
 aip,
 apl,
 amsmath,amssymb,
 reprint,%
superscriptaddress, floatfix
]{revtex4-1}

\usepackage{color}

\usepackage{graphicx}
\usepackage{dcolumn}
\usepackage{bm}

\begin{document}


\title[Sample title]{Thermal detection of single e-h pairs in a biased silicon crystal detector}

\affiliation{Department of Physics, Stanford University, Stanford, CA 94305 USA}
\affiliation{SLAC National Accelerator Laboratory/Kavli Institute for Particle Astrophysics and Cosmology, 2575 Sand Hill Road, Menlo Park, CA 94025 USA}
\affiliation{Department of Physics, Santa Clara University, Santa Clara, CA 95053 USA}
\affiliation{Department of Physics, University of California, Berkeley, CA 94720 USA}

\author{R.K.~Romani} \affiliation{Department of Physics, Stanford University, Stanford, CA 94305 USA}
\author{P.L.~Brink} \affiliation{SLAC National Accelerator Laboratory/Kavli Institute for Particle Astrophysics and Cosmology, 2575 Sand Hill Road, Menlo Park, CA 94025 USA}
\author{B.~Cabrera} \affiliation{Department of Physics, Stanford University, Stanford, CA 94305 USA} \affiliation{SLAC National Accelerator Laboratory/Kavli Institute for Particle Astrophysics and Cosmology, 2575 Sand Hill Road, Menlo Park, CA 94025 USA}
\author{M.~Cherry} \affiliation{SLAC National Accelerator Laboratory/Kavli Institute for Particle Astrophysics and Cosmology, 2575 Sand Hill Road, Menlo Park, CA 94025 USA}
\author{T.~Howarth} \affiliation{Department of Physics, Stanford University, Stanford, CA 94305 USA}\author{N.~Kurinsky} \affiliation{Department of Physics, Stanford University, Stanford, CA 94305 USA} \affiliation{SLAC National Accelerator Laboratory/Kavli Institute for Particle Astrophysics and Cosmology, 2575 Sand Hill Road, Menlo Park, CA 94025 USA}
\author{R.A.~Moffatt} \affiliation{Department of Physics, Stanford University, Stanford, CA 94305 USA}
\author{R.~Partridge} \affiliation{SLAC National Accelerator Laboratory/Kavli Institute for Particle Astrophysics and Cosmology, 2575 Sand Hill Road, Menlo Park, CA 94025 USA}
\author{F.~Ponce} \affiliation{Department of Physics, Stanford University, Stanford, CA 94305 USA}
\author{M.~Pyle} \affiliation{Department of Physics, University of California, Berkeley, CA 94720 USA}
\author{A.~Tomada} \affiliation{SLAC National Accelerator Laboratory/Kavli Institute for Particle Astrophysics and Cosmology, 2575 Sand Hill Road, Menlo Park, CA 94025 USA}
\author{S.~Yellin} \affiliation{Department of Physics, Stanford University, Stanford, CA 94305 USA}
\author{J.J.~Yen} \affiliation{Department of Physics, Stanford University, Stanford, CA 94305 USA}
\author{B.A.~Young} \affiliation{Department of Physics, Santa Clara University, Santa Clara, CA 95053 USA}

\noaffiliation
\date{\today}

\begin{abstract}
We demonstrate that individual electron-hole pairs are resolved in a 1\,cm$^2$ by 4\,mm thick silicon crystal (0.93\,g) operated at $\sim$35\,mK.  One side of the detector is patterned with two quasiparticle-trap-assisted electro-thermal-feedback transition edge sensor (QET) arrays held near ground potential. The other side contains a bias grid with 20\% coverage. Bias potentials up to $\pm$\,160\,V were used in the work reported here. A fiber optic provides 650~nm (1.9\,eV) photons that each produce an electron-hole ($e^{-} h^{+}$) pair in the crystal near the grid. The energy of the drifting charges is measured with a phonon sensor noise $\sigma$\,$\sim$0.09\,$e^{-} h^{+}$ pair. The observed charge quantization is nearly identical for $h^+$'s or $e^-$'s transported across the crystal. 

\end{abstract}


\maketitle

Cryogenic detectors made of ultra-pure single crystals of silicon (Si) and germanium (Ge), and biased with $\sim$100\,V across the crystal, have achieved very low thresholds in the search for dark matter~\cite{CDMSlite2016,CDMSlite2017} 
by converting the ionization signal to a phonon signal with substantially improved resolution over that obtained with a charge amplifier. 
This Neganov-Luke effect \cite{Neganov1985,Luke1990} is proportional to the applied bias voltage across the crystal. The resulting phonon signal is read out using quasiparticle-trap-assisted electro-thermal-feedback transition edge sensors (QETs). The ultimate ionization resolution for these detectors is achieved by counting individual $e^-h^+$ pairs.

Si and Ge are indirect-gap semiconductors, since their conduction band energy minima do not occur at zero momentum. For Si the conduction band has six minima, or valleys, located near the mid points along the [100] direction in the Brillouin zone and produce a highly anisotropic electron mass tensor~\cite{Jacoboni1983,Cabrera2010,Moffatt2016}. As a result, an initially localized group of electrons in a small but uniform electric field will spatially separate into three pairs of clusters along each principle axis, each consisting of electrons occupying one of the three pairs of valleys.~\cite{Moffatt2016,Moffatt2017}
At high temperatures, in high electric fields, or at high impurity concentrations, electrons will undergo frequent quantum transitions between these valleys, resulting in a nearly isotropic electron mobility which is the geometric mean of the nominally anisotropic mobility. 
To model this asymmetry, we developed Monte Carlo simulations of charge transport in our Si and Ge crystals. For the Si detectors in this paper we have isotropic propagation of electrons ($e^{-}$) and holes ($h^{+}$) for electric fields greater than $\sim$\,100\,V/cm, equivalent to a crystal bias of $\sim$40V for the device discussed in this paper.

\begin{figure}[hbtp]
\begin{center}
\includegraphics[height=1.83in]{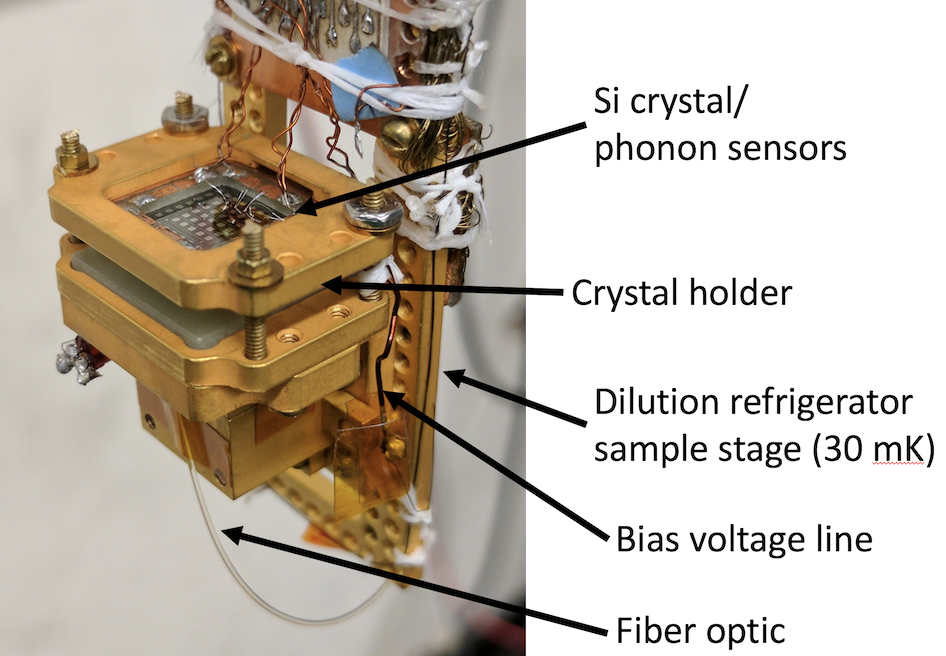}
\caption{\footnotesize (color online)
Photograph of Si detector mounted on mixing chamber stage of KelvinOx 15 dilution refrigerator with phonon sensors on top and bias grid below. A fiber optic illuminates the device from below with 650~nm photons.
}
\label{fig:photo}
\end{center}
\end{figure}

The phonon measurement utilizes the QET~\cite{Irwin1995} advanced athermal phonon sensor technology developed for CDMS\,II~\cite{Akerib_PRD2003} and SuperCDMS~\cite{Pyle2012}.
These sensors are composed of a thick Si (or Ge) crystal patterned using photolithography with aluminum (Al) electrodes connected by tungsten (W). At low temperatures, athermal phonons propagating in the crystal will diffuse into the superconducting Al electrodes on the crystal surface. These phonons are sufficiently energetic to break Cooper pairs in the Al, generating quasiparticles, which diffuse into the thin W film connecting multiple Al electrodes. The W film is operated between the superconducting and normal states as a Transition Edge Sensor (TES) and the excess quasiparticles raise the temperature and resistance of the film.
The TES resistance increase under voltage bias is detected as a decrease in current using SQUID amplifiers.

\begin{figure}[hbtp]
\begin{center}
\includegraphics[width=3.25in]{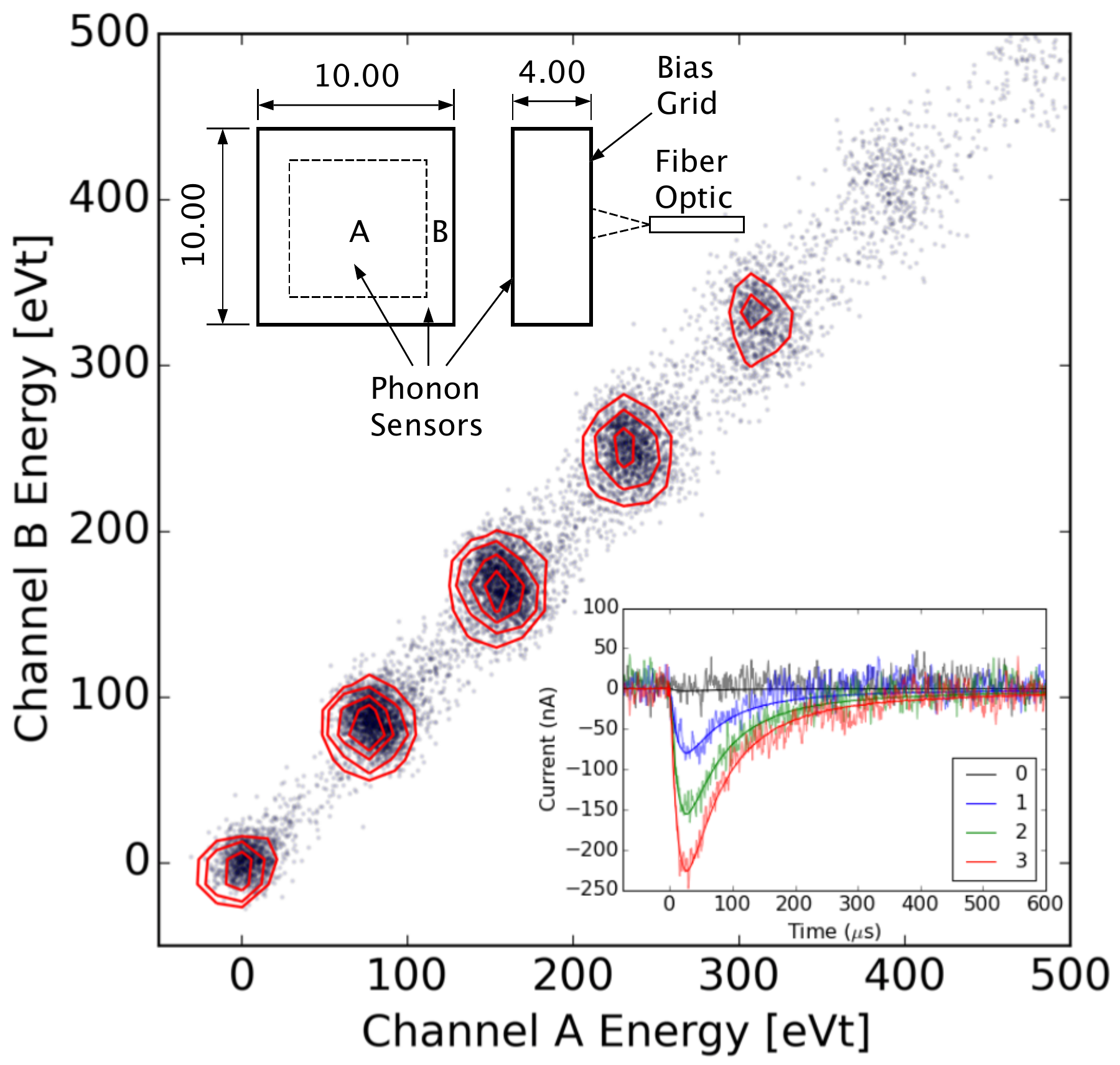}
\caption{\footnotesize (color online)
(upper inset) Schematic of 1\,cm x 1\,cm x 4\,mm thick silicon detector (0.93\,g)  with 160\,V grid bias.  (lower inset) Pulses for 0, 1, 2 and 3 charges per pulse. (plot) Signal energy in inner phonon sensor A versus outer phonon sensor B clearly showing quantization of electrons or holes from photon pulses shining on the grid pattern using a 50\,$\mu$m core fiber optic. For each event the calibrated total phonon energy (eVt) is given by the sum A+B.
}
\label{fig:AvsB}
\end{center}
\end{figure}

A test device was constructed on a 1\,cm x 1\,cm x 4\,mm thick Si crystal (0.93\,g), oriented with the [100] direction perpendicular to the 1\,cm$^2$ face and the side walls along [110]. The front (non-illuminated) face of the crystal was patterned with equal area inner and outer QET phonon channels with a net active Al coverage of 13\% on top of a 40~nm thick amorphous Si layer. Two sensors allow vetoing events near the outer edges if there is significant difference in the phonon collection efficiency.
The back (illuminated) face was patterned with a 20\%  coverage (``parquet pattern") 40\,nm Al electrode overlaid on an 40\,nm aSi film to allow visible photons to be absorbed near the surface of the bulk crystal. The back electrode was biased relative to the front QETs, creating a field across the 4\,mm thick crystal. In operation, this test device was mounted onto the sample stage of a $^3$He\,-$^4$He dilution refrigerator, operating at base temperatures in the range of 30-35\,mK. A photograph of the device mounted on the refrigerator base stage is shown in Fig.\,\ref{fig:photo}.

Electron-hole ($e^{-} h^{+}$) pairs were created in the crystal by illuminating the electrode side with a monochromatic 650\,nm pulsed laser ($\sim$1.91\,eV photons). 
The laser power and pulse width along with optical attenuators control the average number of photons per pulse that reach the sensors. The number of observed photons in any individual pulse is stochastic such that the observed detector response is a convolution of a Gaussian with a Poisson distribution. Setting the average number of photons per pulse to between 1\,-\,10 photons allowed a study of the Neganov-Luke effect for small numbers of charges.

In typical runs, the crystal is cooled to base temperature, and ``neutralized" for $\sim$24-72 hours, where we ground the crystal and illuminate it with the laser at high intensity (2\,mW, 1\,ms, and -10\,dB optical attenuator).
This floods the crystal with $e^-$'s and $h^+$'s, which attach to charge traps. During operation, the Si crystal is biased between -160 and +160\,V and illuminated with the laser at low intensity (200\,$\mu$W, 200\,ns, and -50\,dB optical attenuator). The trace acquisition can be triggered on the laser internal TTL for low noise acquisition or triggered on a threshold to observe the leakage current of the detector. 

For the crystal biased at 160\,V, the laser instensity averaging $\sim$2 photons per pulse and the acquisition system triggering on the laser TTL, a comparison of the total collected energy in eV (eVt) in each QET channel is shown in Fig.\,\ref{fig:AvsB}. The contours show that channel noise is uncorrelated and that the channels measure comparable energies for a given laser pulse. The amplitude of the acquired traces was estimated using a matched filter, with fits shown in the inset of Fig.\,\ref{fig:AvsB}. These data demonstrate a highly precise measurement of the quantized $e^{-}h^{+}$ pair peaks generated by the laser. 

These data have a quadratic nonlinearity for $e^{-}h^{+}$ peak position versus amplitude, which we correct using $a_{linear}=a[1+0.016a/(160\,\mathrm{eVt})]$. This small effect is due to the series resistance in the bias circuit, which prevents purely linear electrothermal feedback.~\cite{Shank2014}
We calculate the energy collection efficiency of the device by comparing the inferred energy absorbed from the current change to the absolute energy calibration from the laser. When phonon energy is absorbed, the current through the TES decreases, producing a decrease in current (and therefore a decrease in Joule heating) proportional to the energy absorbed. For our sharp TES transition near 51\,mK, this energy input changes the resistance of the device, but leaves temperature largely unchanged, so that the increase in energy input is balanced by the decrease in Joule heating, thus allowing us to find the absorbed power as
\begin{align}
E_{abs}&\approx -\int_{0}^{T}\Delta P_J(t)dt \\
&\approx \left(2R_lI_0-V_b\right)\int_{0}^{T}\Delta I dt +R_l\int_{0}^{T}(\Delta I)^2dt
\end{align}
where $R_l$ is the resistance in series with the TES in the voltage-biased topology, $V_b$ is the voltage bias, $I_0$ is the TES bias current, and we have assumed a sharp TES transition to simplify this expression.  We find that the single $e^-h^+$ peak shown in Fig.\,2 (for summed A+B), 
with 161.9\,eVt from 160\,eV Luke gain plus 1.9\,eV of the original photon, corresponds to $E_{abs}\,\sim\,8$\,eV, giving a measured efficiency of $5\pm 1$\%.  The systematic uncertainty is due to uncertainties in bias circuit components and operating point resistances.  

\begin{figure}[hbtp]
\begin{center}
\includegraphics[width=3.2in]{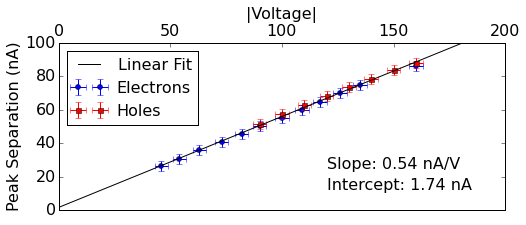}
\includegraphics[width=3.2in]{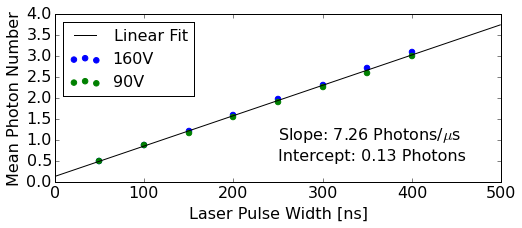}
\caption{\footnotesize (color online)
(top) Linearity of phonon amplitude versus bias voltage across the crystal.  (bottom) Linearity of phonon energy versus photon number per laser pulse.
}
\label{fig:linearity}
\end{center}
\end{figure}

Figure\,\ref{fig:linearity} (top) shows the position of the first $e^{-}/h^{+}$ pair peak across a range of voltages where quantization is detectable for both $e^{-}$ and $h^{+}$ propagation (positive and negative bias).
The linearity with voltage demonstrates that the athermal phonon collection efficiency for Neganov-Luke phonons is independent of both E-field strength and excitation type throughout this range of biases.

\begin{figure}[hbtp]
\begin{center}
\includegraphics[width=3.25in]{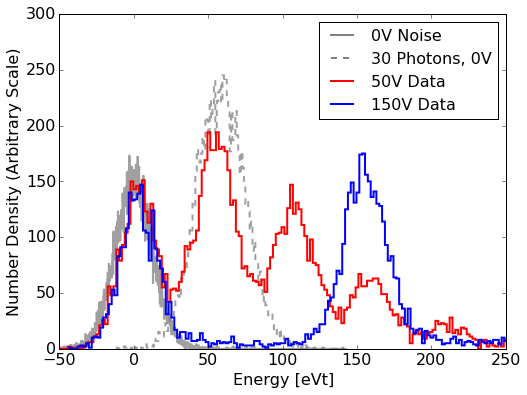}
\caption{\footnotesize (color online)
Four superimposed histograms with total phonon energy scale (summed A+B in eVt) showing (1) the noise peak with no laser pulses and grounded crystal, (2) 30 average photons per laser pulse with crystal grounded, and calibration of phonon energy using small resolved charge across the crystal with (3) 50\,V and (4) 150\,V biases across the crystal (agreement to $\sim$5\% - see text).
}
\label{fig:zero_bias}
\end{center}
\end{figure}

This linearity, coupled with the invariance of the phonon noise with voltage (as demonstrated in Fig.\,\ref{fig:zero_bias}), means that as long as this trend continues to higher voltages, we can continue to expect linear gains in signal to noise. In addition, we show that the relationship between mean photon number and laser power is also highly linear and invariant to voltage, allowing us to compare the high voltage and 0\,V energy distributions to check the absolute phonon energy calibration.

Figure\,\ref{fig:zero_bias} demonstrates this 0\,V calibration, utilizing the measured photon yield as a function of laser power shown in Fig.\,\ref{fig:linearity} (bottom) to compare the signal from a 30 photon pulse to the high-voltage data. These 0\,V pulses should deposit 1.91\,eV per photon or 57\,eV into the phonon system (assuming the recovery of all of the gap energy as $e^{-}$ and $h^{+}$ recombine at surfaces). Comparison with the first photon peak at 50\,V bias which should produce 52\,eV of phonons or $1/3$ of the first photon peak at 150\,V bias shows agreement is good to $\sim$5\%, within possible systematics in the zero bias measurement from any residual space charge that would add Neganov-Luke phonon energy or local trapping of $e^{-}$'s or $h^{+}$'s which would prevent the gap energy from returning to the phonon system. This good agreement suggests that neither effect is significant.

\begin{figure}[hbtp]
\begin{center}
\includegraphics[width=3.25in]{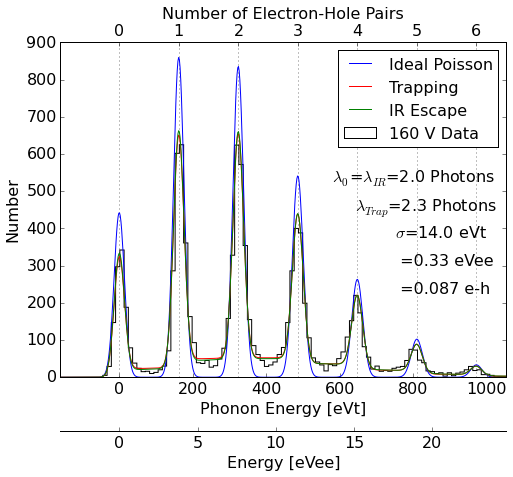}
\caption{\footnotesize (color online)
Histogram of summed A+B data from Fig.\,\ref{fig:AvsB}  showing the excellent fit for a Poisson distribution. A small nonlinearity is taken out of the data prior to the fits (see text). The integer number of $e^{-} h^{+}$ pairs is shown above, the phonon crystal energy below (eVt), and an electron-equivalent energy scale (eVee) at bottom using the standard 3.8\,eV per $e^{-} h^{+}$ pair.~\cite{Vavilov1962} Fits are performed including trapping and impact ionization  (see text).}
\label{fig:quantization}
\end{center}
\vspace{-15pt}
\end{figure}

Figure \ref{fig:quantization} shows that $\sim$15\% of the events are distributed in between the quantized photon peaks. For these events, one or more of the produced ionized excitations did not traverse the entire crystal, and thus it’s Neganov-Luke phonon production was incomplete, and non-quantized. This can occur if an ionized excitation was trapped in the bulk while drifting.  A second possibility is impact ionization, where a drifting excitation scatters off an occupied impurity state releasing a non-paired excitation that drifts across only a fraction of the crystal.  Finally, subgap photons produced in coincidence when the laser is pulsed could also interact with filled impurity states, again producing an unpaired excitation.

These three models were fit to the data by a maximum-likelihood fit in which the noise variance, peak separation, mean photon number, impact ionization probability, ionization trapping probability, and the average sub-gap IR absorption number were allowed to vary. Impact ionization events contribute less than $\sim$1/5 of the fill-in events, which are $\sim$15\% of all events in the data.  Impact ionization is not a major contribution since it cannot produce any of the events seen between the 0 and 1 $e^-h^+$ peaks. The best-fit models for a typical dataset at 160\,V can be seen in Fig.\,\ref{fig:quantization} compared to the ideal Poisson model with the probability of these secondary processes set to 0.  When there is a contribution from only one of the three models, both subgap IR photon absorption and ionization trapping produce good fits.

These fits were performed both as a function of bias voltage and laser input power. A striking trend is that the secondary process probability (the total amount of fill-in between the peaks) is found to be independent of voltage, 
which disfavors the trapping model since we expect the trapping length should be a function of the mean field strength in the crystal.~\cite{Sundqvist2012}

In the subgap infrared absorption model, fitting data with different average photon number to the same model requires that the sub-gap IR photon flux be proportional to the average number of above gap photons (proportional to the pulse time of the laser). This is certainly quite reasonable and perhaps even expected. In the future, we will modify our fiber setup to be a single-mode fiber instrumented with IR filters to attempt to substantially suppress this probable background. These upgrades will have the added benefit of further isolating the detector from room-temperature IR not coincident with the laser pulse, which may dominate the leakage rate.

This letter demonstrates the operation of the first phonon-based detector capable of resolving single charges. This detector has a demonstrated resolution of $\sim$0.09 $e^{-} h^{+}$ pairs and a fiducial mass of 0.93\,g, and represents a new generation of gram-scale detectors capable of measuring single energy deposits on the order of the Si bandgap ($\sim$1.2\,eV) in real-time with significantly better pileup-rejection ($\sim$10\,$\mu$s) and larger mass than existing CCD-based technologies. We have also shown that this device continues to operate linearly across the range of input conditions that we have tested; suggesting that the signal/noise may continue to improve with increased bias voltage. In addition, a device with two-sided phonon readout and higher collection efficiency, may improve the timing, energy efficiency and resolution by a factor of $\sim$3.

In the short term, such devices will allow the SuperCDMS collaboration to measure not only the average ionization yield of nuclear recoils down to the production of the first $e^{-} h^{+}$ pair
in Si and Ge, but also measure the full probability distribution of the nuclear recoil ionization yield as a function of recoil energy - essential for nuclear recoil direct detection dark matter searches based on ionization measurement in the $\mathrm{100\,MeV < M_{DM}< 6\,GeV}$. In addition, as shown in this paper, the spectral information gained with quantization allows a better understanding of the physics of athermal phonon detectors using Neganov-Luke amplification, such as kg-scale SuperCDMS SNOLAB detectors.

In the longer term such quantization could be used to distinguish between the background of electron recoils from ambient radioactivity and low energy nuclear recoils. Such nuclear recoils are inefficient at producing $e^-h^+$ pairs, requiring roughly an order of magnitude more recoil energy deposition than the 3.8 eV per pair for electron recoils.\cite{Chavarria2016}  That extra phonon energy added to the Neganov-Luke amplification increases the total crystal energy (eVt) of nuclear recoils over those for electron recoils with the same number of $e^-h^+$ pairs, thereby placing such events before and between the first few electron recoil peaks.
This capability could enable a future upgrade to the SuperCDMS SNOLAB experiment with sensitivity to the solar neutrino floor, as well as precision experiments to probe coherent scattering of neutrinos from nuclei.\cite{Akimov2017}  Finally, such quantization could allow a direct detection experiment to differentiate between a hypothetical very light dark matter candidate that interacts electronically from a higher mass dark matter candidate that scatters off a nucleus but produces similar ionization.

This work was supported in part by the U.S. Department of Energy and by the National Science Foundation. The authors are also especially grateful to the staff of the Varian Machine Shop at Stanford University for their assistance in machining the parts used in this experiment.


\nocite{*}

\end{document}